\documentstyle[aps, epsfig, float, preprint ]{revtex}
\tightenlines
\begin{document}
\title{Electronic properties and Fermi surfaces\\\ MgCNi$_{3}$ and related intermetallics}

\author{I.\,R.\,Shein\/\thanks{e-mail:irshein@mail.ur.ru}, N.\,I.\,Medvedeva, A.\,L.\,Ivanovskii}

\address{Institute of Solid State Chemistry, Ural Branch of the Russian
Academy of Sciences, 620219 Ekaterinburg, Russia}

\maketitle
\begin{abstract}
The band structure of the new perovskite-like superconductor
MgCNi$_{3}$ was studied by the self-consistent FP-LMTO method. The
superconducting properties of MgCNi$_{3}$ are associated with an
intensive peak in the density of Ni3d states near the Fermi level.
The absence of superconductivity for nonstoichiometric
compositions MgC$_{1-x}$Ni$_{3}$ is due to the transition of the
system to the magnetic state. The possibility of superconductivity
was discussed for intermetallics ScBNi$_{3}$, InBNi$_{3}$,
MgCCo$_{3}$ and MgCCu$_{3}$ which are isostructural with
MgCNi$_{3}$.

\end{abstract}


\newpage

The discovery (2001 [1]) of the superconducting transition
(T$_{c}\approx$ 39 K)for intermetallic MgB$_{2}$ gave impetus to
wide search for novel superconductors (SC) among related compounds
which is carried out now in three basic directions. The first
direction is to the expansion of the class of MgB$_{2}$-based SC
by doping it or by creating superstructures [2]. In the second
direction SC candidates are looked for among binary or
multicomponent phases with elements of structural or chemical
similarity with MgB$_{2}$. As a result, critical transitions were
found for ZrB$_{2}$ (5.5 K [3]), TaB$_{2}$ (9.5 K [4]), Re$_{3}B$
(4.7 K [5]), the new beryllium boride phase (0.72 K, the
composition BeB$_{2.75}$, the unit cell containing 110.5 atoms
[6]). The development of the third direction was initiated by the
observation [7] of superconductivity in the ternary intermetallic,
viz. the perovskite-like MgCNi$_{3}$ (T$_{c} \approx$ 8K). Some
circumstances make the finding [7] especially interesting.\\\
1. The high-symmetry structure of MgCNi$_{3}$ (space group Pm3m)
is a favourable factor for superconductivity. However, all known
to date SC perovskites contain oxygen atoms in 3C-type sites (0;
1/2; 1/2), the electron-hole states of which play the basic role
in the formation of superconductivity [8]. For MgCNi$_{3}$, those
sites are occupied by Ni atoms, i.e. the mechanism of SC should be
of a fundamentally different nature.\\\
2. The majority of known "non-oxide" perovskite-like phases
MXM$^{'}_{3}$ (the so-called antiperovskites, where M = Zn, Al,
Ga, In, Sn; M' = Mn, Fe; X = C, N) exhibit ferro-,
antiferromagnetic properties, or have a more complicated spin
ordering [9]. Hence, in the series of structural analogs,
MgCNi$_{3}$ can be considered as a "boundary" phase between
perovskite-like SC (oxides) and magnetics (non-oxygen perovskites).\\\
3. The closest chemical analogs of MgCNi$_{3}$ superconducting
borocarbides of inermetallics (BCI) of the general composition
LnM$_{2}$B$_{2}$C. They also include nickel-containing phases
LuNi$_{2}$B$_{2}$C (T$_{c} \approx$ 16K), YNi$_{2}$B$_{2}$C
(T$_{c} \approx$ 15,6K). However, as distinct from MgCNi$_{3}$,
BCI (i) are magnetic superconductors, (ii) they have a
quasi-two-dimensional structure composed of (Lu,Y)C layers and
NiB$_{4}$ tetrahedra [10], and (iii) the content of Ni (magnetic
metal) in BCI is considerably smaller (35.6 - 48.9 at.\%) than
in MgCNi$_{3}$ (82.9 at.\%).\\\
The first investigations of some properties of MgCNi$_{3}$, namely
the critical field (H$_{c2}$), Hall coefficient, other
electrophysical characteristics [11-14], made it possible to
assign MgCNi$_{3}$ to "conventional" superconductors of the II-nd
type with an electron-phonon type of interactions. In that case
the critical temperature can be estimated from the McMillan
formula T$_{c} \sim <$w$>$exp(f($\lambda$)), where  $<$ w $>$ is
an averaged phonon frequency, $\lambda$ is the electron-phonon
interaction constant $\lambda \sim$ N($E_{F}$)$<$$I^{2}$$>$, where
N($E_{F}$) is the density of states on a Fermi level and $<I^{2}>$
is average square of a matrix device an electron - phonon
interaction. Therefore the information on the band structure plays
a significant role both for the interpretation of superconducting
(and other) properties of MgCNi$_{3}$ and for the quest of possible SC analogs.\\\
In this paper we present the results of investigations of the band
structure of the new SC MgCNi$_{3}$, discuss the effect of C
vacancies (nonstoichiometry in carbon sublattice) on its
electronic properties and analyze the peculiarities of electronic
and magnetic states for related perovskite-like alloys
(ScBNi$_{3}$, InBNi$_{3}$, MgBCo$_{3}$ and MgCCu$_{3}$) as
probable superconductors. The self-consistent spin-polarized
full-potential linear muffin-tin orbital (FP-LMTO) method [15] in
the framework of local (spin) density approximation (LDA) with
relativistic effects according to the scheme [16] and with an
exchange-correlation potential proposed in [17] was used in the
calculations. In the structure of MgCNi$_{3}$, atoms occupy the
following positions: 3Ni (0; 1/2; 1/2), Mg (0; 0; 0), C (1/2; 1/2;
1/2). Their coordination polyhedra (CP) are [NiC$_{2}$Mg$_{4}$]
and [CNi$_{6}$] octahedra for Ni and C and the cubooctahedron
[MgNi$_{12}$] for magnesium. The theoretical equilibrium lattice
parameter of MgCNi$_{3}$ (3.721 $\AA$) defined by minimization of
the total energy appeared to be in good agreement with the
experimental value 3.8066 $\AA$ (for MgC$_{0.96}$Ni$_{3}$ with T = 0 K) [18]. \\\
The band structure for MgCNi$_{3}$ (Figs. 1-3) are a good
agreement with the calculation of authors [19]. The most important
feature of the MgCNi$_{3}$ spectrum is the presence of an
intensive DOS peak near the E$_{F}$ associated with the
quasiplanar $\pi$-antibonding Ni3d bands (in the directions X-M
and M-$\Gamma$ of the Brillouin zone), see Fig. 1. The N(E$_{F})$
is located on the high-energy slope of the peak. The value of
N(E$_{F})$ is 4.57 states/eV, which accords well the
full-potential FLAPW calculation (4.99 states/eV [20]). The
expansion of N(E$_{F}$) into orbital components (N$_{l}(E_{F})$)
shows (Table 1) that the maximum contribution to N($E_{F}$) (4,04
state/eV, or 88,2\%) is due to the Ni3d states. The contributions
of C2s,p and Mg3s,p,d states to N($E_{F}$) are 0.23 (5.08 \%) and
0.14 states/eV (3.03 \%) respectively. The Stoner parameter S =
N(E$_{F}$)I$_{ex}$ (I$_{ex}$ - exchange integral) is $\approx$
0.55, the magnetic moments on atoms are absent. The upper of the
two antibonding Ni3d bands, which has a more pronounced
dispersion, is responsible for the electronic type of the Fermi
surface (Fig. 3) having the form of of spheroids near the $\Gamma$
point and small sheets along the boundaries and angles of the
Brillouin zone. The flatter Ni3d band produces clover-type sheets
centered in the X point and cigar-like figures along the
$\Gamma$-R direction.\\\
To compare individual interatomic bonds we calculated used by the
tight-binding band method the crystal orbital overlap populations
(COOP) of MgCNi$_{3}$ and ScBNi$_{3}$. The corresponding values
for MgCNi$_{3}$ were 0.298 (Ni-C), 0.027 (Ni-Ni) and 0.039 e/bond
(Ni-Mg). So, the Ni-C bonds (in the [CNi$_{6}$] CP, Fig. 4) are
the dominant interatomic interactions in MgCNi$_{3}$. The C-Mg
bonding is negligibly small (0.002 e/bond). For ScBNi$_{3}$, these
values are 0.338 (Sc-B), 0.050 (Ni-Ni), 0.033 (Ni-Sc) and 0.005
e/bond (B-Sc). These results make it possible to explain the data
[13] on temperature dependence of the Debye-Waller factor (DWF)
for atoms in MgCNi$_{3}$. The minimum (isotropic) temperature
factor of carbon corresponds to its most bounding (and
high-symmetry - in the center of Ni$_{6}$ octahedron) state in the
crystal, whereas for Ni the DWF has a greater value and is
anisotropic: in the CP of Ni (NiC$_{2}$Mg$_{4}$ octahedra) the
COOP of different-type (Ni-C and Ni-Mg) bonds differ by an order
of magnitude. The observed minimum mean-square displacements
($U_{11}$) of nickel correspond to the directions of the strongest
Ni-C bonds. \\\
Based on Fig. 2, in the framework of the rigid-band model it can
be expected that the introduction of electronic or hole dopants
into MgCNi$_{3}$ will bring about an increase or decrease in
N(E$_{F})$ respectively. In the first case, the SC of the system
is expected to worsen. Doping by holes, which promotes an increase
in N(E$_{F}$) and is a favorable factor for a rise in T$_{c}$,
can, however, cause a transition of the system to the magnetic
state with loss of SC. A similar band structure (an intensive
near-Fermi peak of metallic states generally indicating the
instability of the nonmagnetic state of the system) is realized
for the superconducting BCI and determines the formation of atomic
magnetic moments (MM) [4].\\\
We have carried out the calculations of the systems simulating the
these modifications. The effect of a decrease in the occupation of
energy bands was viewed as a result of the presence of vacancies
in the C-sublattice of MgCNi$_{3}$ (hypothetical perovskite
MgENi$_{3}$ with "empty" C sublattice, E - vacancy) or the Ni -$>$
Co substitution (MgCCo$_{3}$). Electron concentration growth was
simulated using the MgCCu$_{3}$ phase as an example. In addition,
stable boron-containing phases isostructural and isoelectronic
with MgCNi$_{3}$, namely ScBNi$_{3}$ and InBNi$_{3}$ with lattice
parameters given in [21], were considered for the first time as
possible SC. For the "nonstoichiometric" antiperovskite it was
found that MgENi$_{3}$ is in the magnetic state, atomic MM being
0.44 and -0,05 $\mu$B for Mg and Ni atoms respectively. An
analogous result was obtained also for InENi$_{3}$: MM are 0.20
$\mu$B for Ni and -0.01 $\mu$B for In atoms. Hence, the
experimentally noticed [7] condition of obtaining the
superconducting MgCNi$_{3}$ as a phase of strictly
stoichiometrical composition MgC$_{1-x}$Ni$_{3}$ samples (x $>$
0,1) lose SC) is primarily by the peculiarities of its electronic structure.\\\
The ground state of the antiperovskite MgCCo$_{3}$ appears from
calculations to be magnetic, MM of atoms being 0.36 for Co and
-0.05 $\mu$B for Mg. For MgCCu$_{3}$, we found that (i) the
electron concentration growth results in the filling of
antibonding bands, Fig. 1, and (ii) N(E$_{F})$ with dominating
delocalized sp states decreases by more than an order as compared
to MgCNi$_{3}$, Fig. 3. The data obtained allow us to explain the
changes in the superconducting properties of MgCNi$_{3}$ when the
Ni sublattice is doped with 3d metals. It is known that the
critical transition temperature for MgCNi$_{3-x}$M$_{x}$ alloys (M
= Mn, Co, Cu) [13, 14] drops if (i) the concentration of dopants
increases or (ii) their atomic number diminishes (Co -$>$ Mn). It
is seen from the dependence between $T_{c}$ and the concentration
of copper that $T_{c}$ decreases progressive in the limits 0 $<$ x
$<$ 0.1. Partial substitution of cobalt atoms for nickel leads to
disappearance of superconductivity already for x=0.03. The effect
of superconductivity suppression strengthens for Ni -$>$ Mn
substitution. According to calculated data, the nature of the
above effect differs essentially for different dopants. For alloys
with substituted Mn and Co, this is due to spin fluctuations; as
for Cu, the reason is with the growth of the total electron
concentration in the system and an abrupt decrease in N(E$_{F})$.
The results of calculations of boron-containing antiperovskites
ScBNi$_{3}$, InBNi$_{3}$ are given in Figs. 1-4 and Table 1.
Charge density for MgCNi$_{3}$ and $ScBNi_{3}$ (Fig. 4) show that
covalent overlapping of Sc-B bonds is stronger than of Mg-C bonds,
which completely coincides with the presented above data obtained
by us in the framework of the tight-biding band theory.\\\
In going from MgCNi$_{3}$, ScBNi$_{3}$ and InBNi$_{3}$ the Fermi
level shifts to the high-energy region of the Ni3d peak, and
N(E$_{F})$ decreases essentially. In this series of compounds, one
of the antibonding bands lowers relative to the Fermi level along
the M-$\Gamma$ direction and shifts upwards in the point X. In the
M point, it shifts upwards, and in the direction $\Gamma$-R goes
below the Fermi level. These changes in the band structure lead to
the modification of the Fermi surface. Electronic-type sphere-like
surfaces near the $\Gamma$ point undergo insignificant changes.
The quasi-cylindric electronic surface along the BZ edges
increases in this series. No hole cigar-like figures along the
$\Gamma$-R direction are present for ScBNi$_{3}$ and InBNi$_{3}$,
and clover-type sheets on the faces of the BZ centered in the X
point increase for ScBNi$_{3}$ and degenerate in a spheroid for
InBNi$_{3}$. Thus, the Fermi surface topology for the considered
compounds preserves the basic features of the superconducting
MgCNi$_{3}$. Fig. 3 also displays the Fermi surface for
MgCCu$_{3}$. For this compound, the shift of the Fermi level
results in a qualitatively different topology: neither electronic-
nor hole-type surfaces are present in the vicinity of
points $\Gamma$ and X.\\\
This double substitution causes changes in the electronic
structure of the above isoelectronic systems, which may be
expected to exhibit superconducting properties if doped with a
small quantity of holes. The most probable ways are introduction
of B vacancies (nonstoichiometry in boron sublattice, especially
as for InBNi$_{3}$ this possibility is known from experiment [21])
or partial replacement of Sc by I, II group atoms. Doping of the
Ni sublattice with magnetic impurities (Co, Mn) can be more
problematic because of magnetic instability.\\\
Thus, the band structure of the new perovskite-like superconductor
MgCNi$_{3}$ has been examined. Its superconducting properties are
associated with the presence of the intensive Ni3d DOS peak near
the Fermi level. Impairment of SC characteristics of MgCNi$_{3}$
doped with holes MgC$_{1-x}$Ni$_{3}$ compositions or with Co, Mn,
is explained by the transition of the system to the magnetic
state. The impairment of SC as a result of electron doping (alloys
MgC$_{1-x}$Cu$_{3}$) is due to the filling of antibonding states
and the abrupt decrease in N(E$_{F})$. It is pointed out that
stoichiometric antiperovskites ScBNi$_{3}$ and InBNi$_{3}$ may
exhibit superconducting properties.

\center{\textbf{{FIGURES:}}}
\par
 \center{\textbf{{Fig.1. Energy bands 1 - MgCNi$_{3}$, 2 - InBNi$_{3}$, 3 - ScBNi$_{3}$, 4 -
MgCCu$_{3}$.}}}

\center{\textbf{{Fig.2. Density of state 1 - MgCNi$_{3}$, 2 -
ScBNi$_{3}$, 3 - InBNi$_{3}$ : I - Total DOS, II - atoms B,C DOS,
III - atom Ni DOS, IV - atom Mg, Sc, In DOS}}}

\center{\textbf{{Fig.3. Fermi surfaces for: 1- MgCNi$_{3}$, 2 -
InBNi$_{3}$, 3 - ScBNi$_{3}$, 4 - MgCCu$_{3}$.}}}

\center{\textbf{Fig.4. Charge densities in 1- MgCNi$_{3}$ and 2 -
ScBNi$_{3}$ in 1/eV$^{3}$.}}


\begin{thebibliography}{21}

\bibitem{Nagamatsu}
J. Nagamatsu, N. Nakagawa, T. Muranaka, Y. Zenitani ets., J.
Akimitsu Nature, {\bf 410}, 63 (2001).

\bibitem{Medvedeva}
N.I.Medvedeva,U.E.Medvedeva, A.L.Ivanovskii etc., Letter in JETP,
{\bf 73}, 378 (2001).

\bibitem{Gasparov}
V.A.Gasparov, N.S.Sidorov and M.P. Kulakov, cond-mat/0104323
(2001).

\bibitem{Kaczorowski}
D.Kaczorowski, J.Klamut and A.Zaleski, cond-mat/0104479 (2001).

\bibitem{Strukova}
G.K. Strukova, V.F.Degtyareva, D.V. Shivkun ets., cond-mat/0105293
(2001).

\bibitem{Young}
D.Young, P. Adams, J. Chan ets.,cond-mat/0104063 (2001).

\bibitem{He_Huang}
T. He, Q. Huang, A.P. Ramirez ets., cond-mat/0103296 (2001).

\bibitem{Taraphder}
A. Taraphder, R. Pandit, H.R. Krishnamurthy and T.V. Ramakrishnan,
Int. J. Modern Phys., {\bf B10}, 863 (1998).

\bibitem{Ivanovskii001}
A.L. Ivanovskii, Successes of chemistry (Rus), {\bf64}, 499
(1995).

\bibitem{Ivanovskii002}
A.L. Ivanovskii, Successes of chemistry (Rus), {\bf67}, 493
(1998).

\bibitem{Li_Fan}
S.Y. Li, R. Fan, X.H. Chen ets., cond-mat/0104554 (2001).

\bibitem{Mao}
Z.Q. Mao, M.M. Rosario, R. Nelson ets., cond-mat/0105280 (2001).

\bibitem{Hayward}
M.A. Hayward, M.K. Haas, T. He ets., cond-mat/0104541 (2001).

\bibitem{Ren}
Z.A. Ren, G.C. Che, S.L. Jia ets., cond-mat/0105366 (2001).

\bibitem{Methfessel}
Methfessel M., Scheffler M. Physica B., {\bf 172Ð},175 (1991).

\bibitem{S.Savrasov001}
Savrasov S.Y. Phys. Rev., {\bf B54}, 16470 (1996).

\bibitem{Vosko}
Vosko S.H., Wilk L., Nusair M. Canadian J. Phys., {\bf58}, 1200
(1980).

\bibitem{Huang_He}
Q. Huang, T. He, K.A. Regan ets., cond-mat/0105240 (2001).

\bibitem{Korea}
J.H. Shim and B.I. Min. cond-mat/0105418 (2001).

\bibitem{Singh}
J.D. Singh and I.I. Mazin. cond-mat/0105577 (2001).

\bibitem{Kusma}
A.P. Kusma, Crystal chemistry of borides (Rus), L'vov, "Vysha
shkola", 1983.

\end{thebibliography}
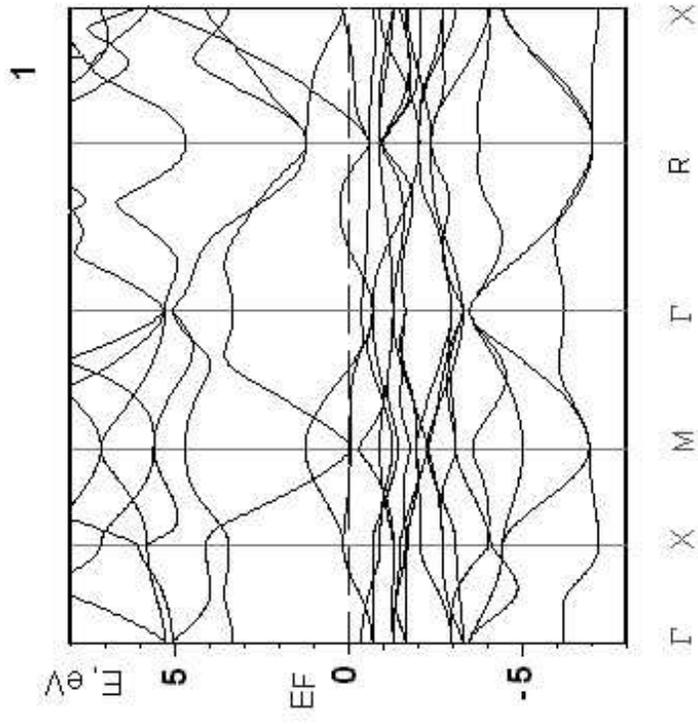
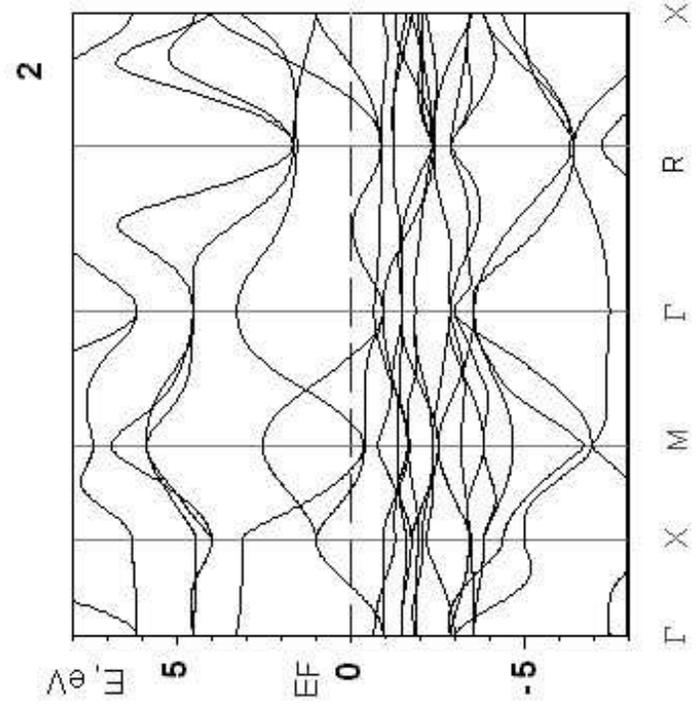
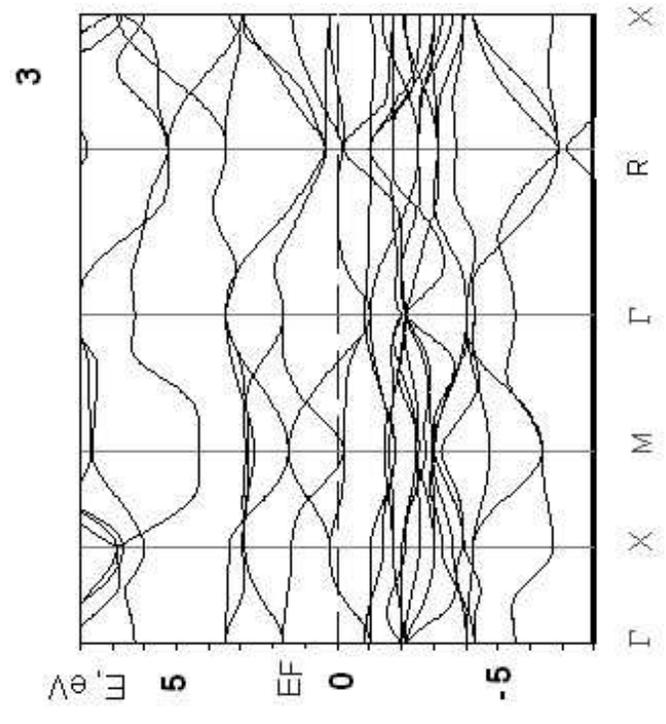
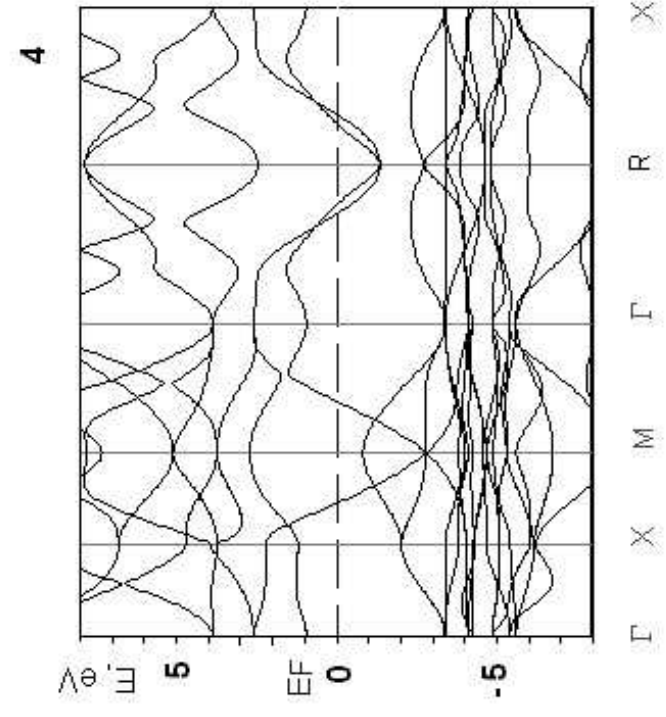
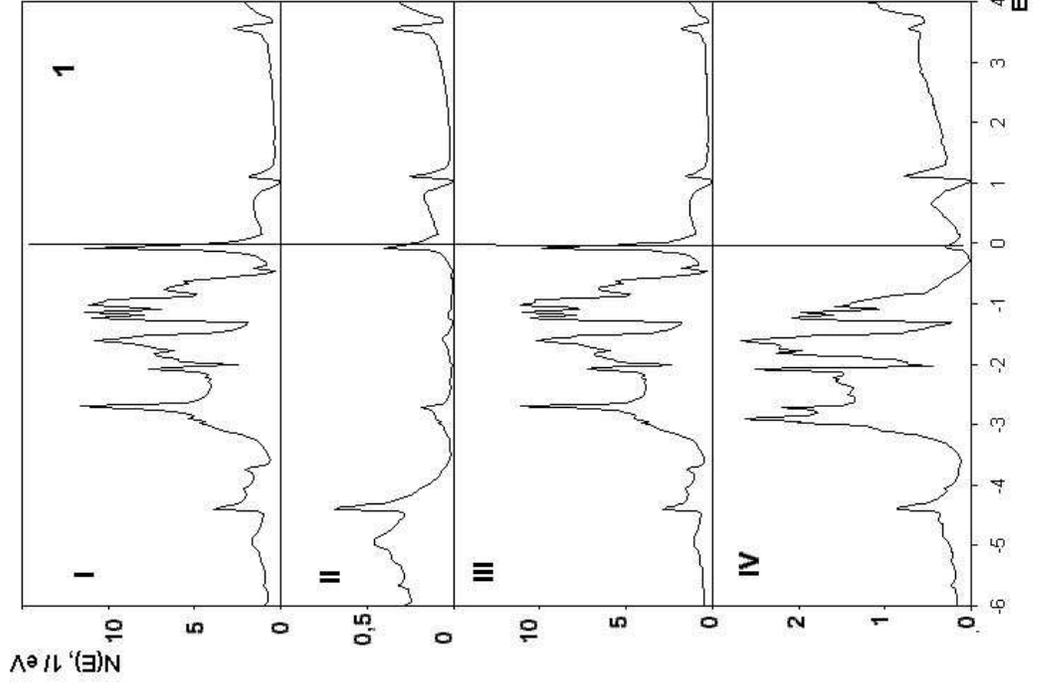
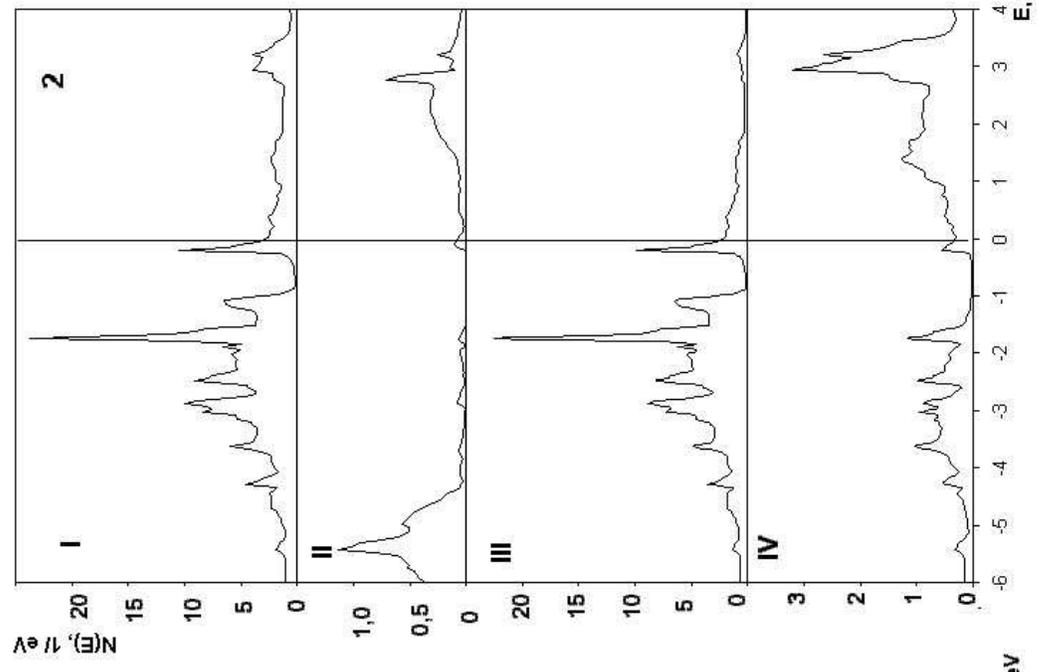
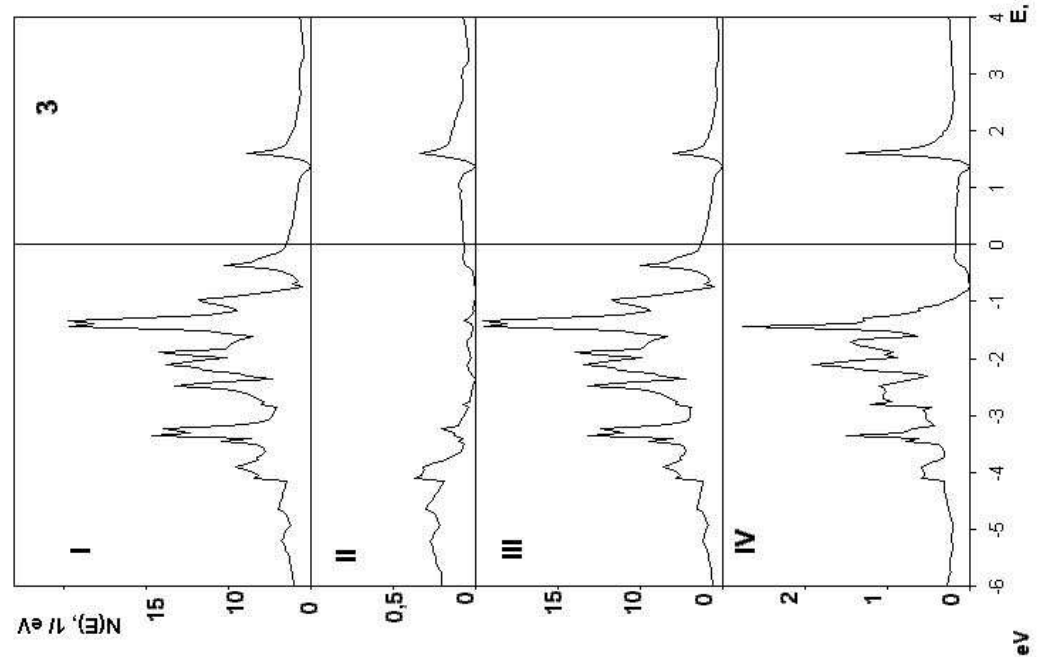
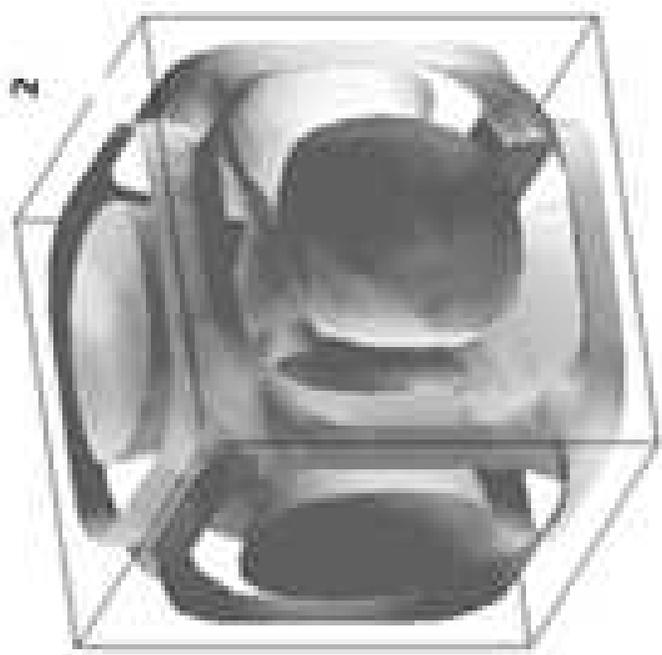
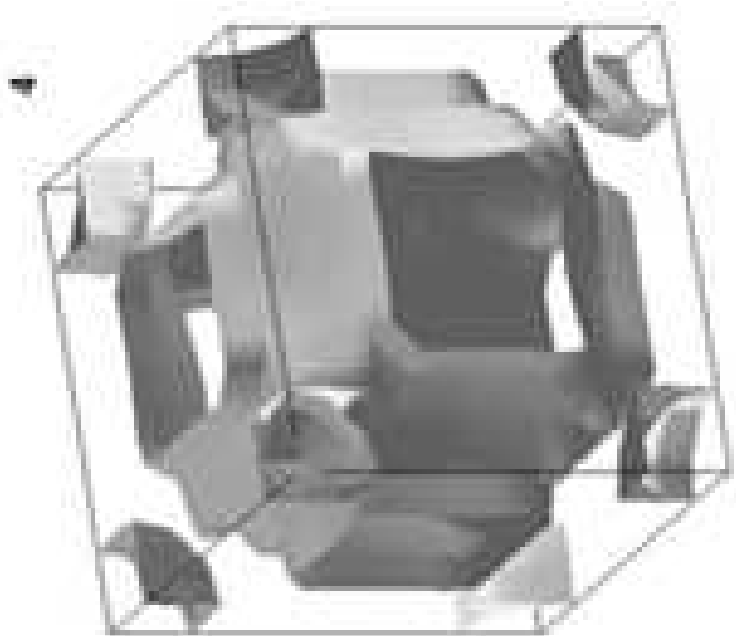
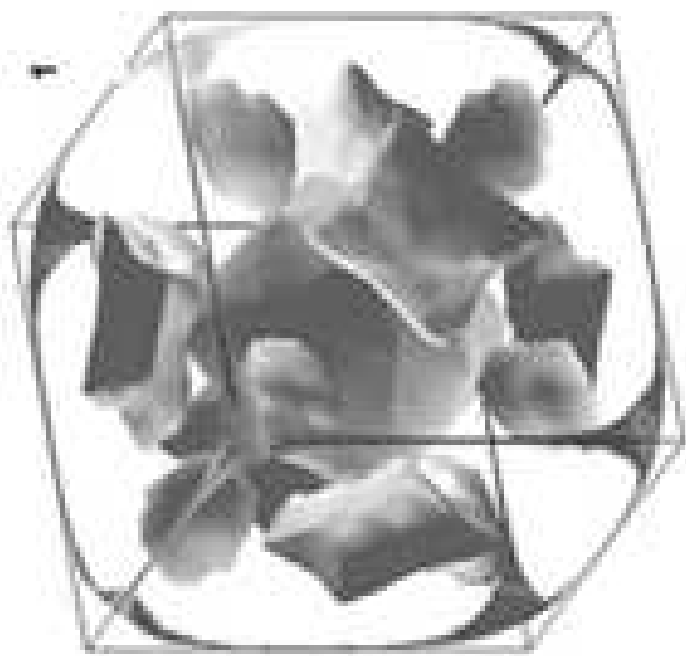
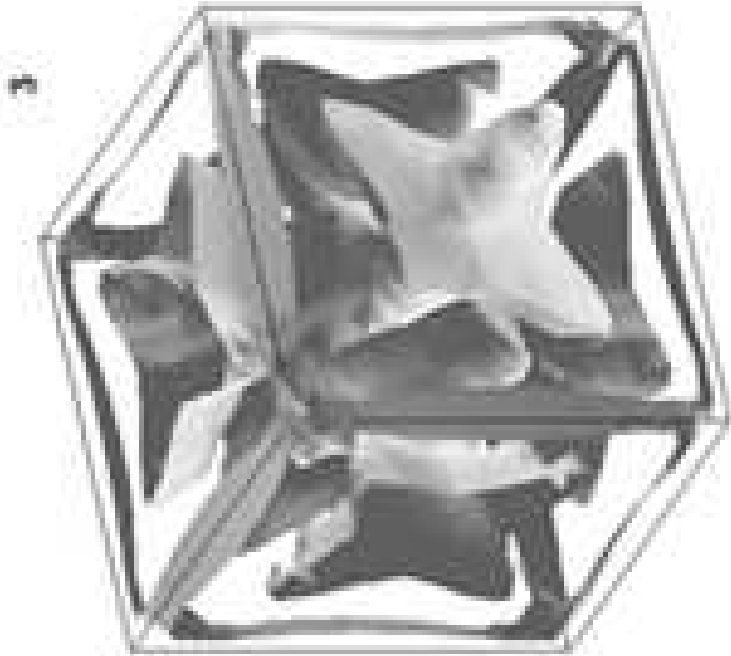
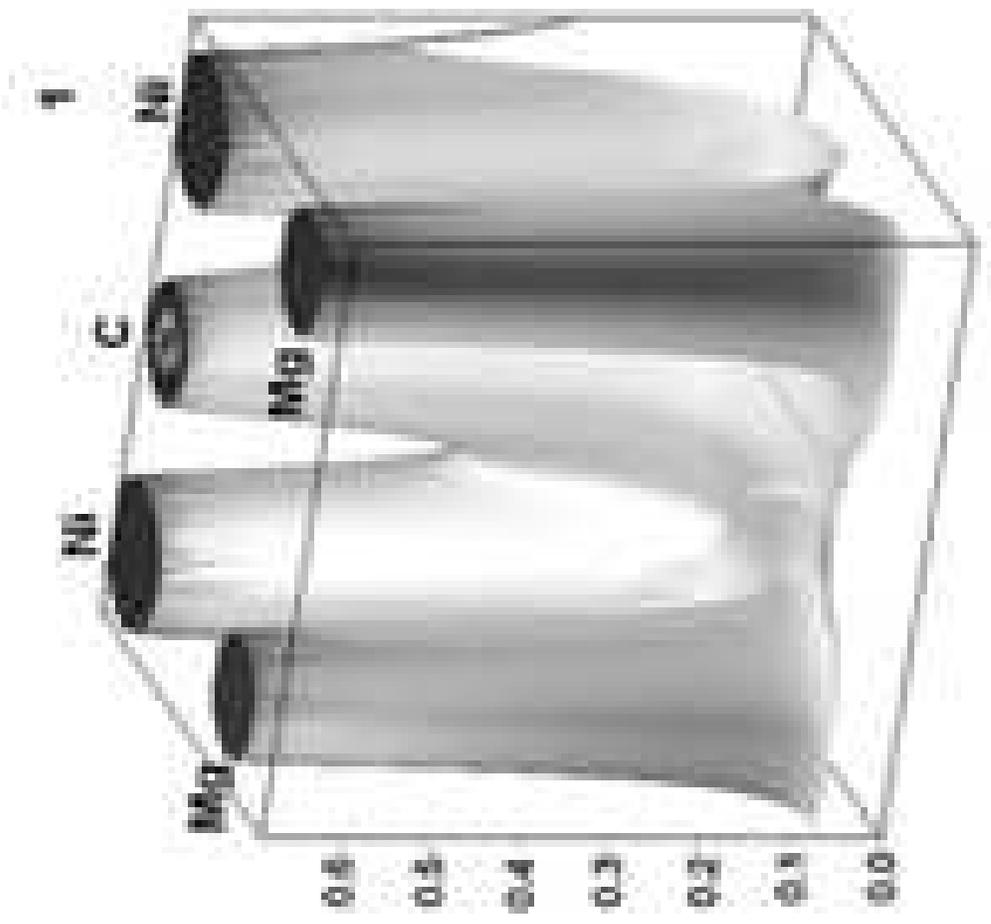
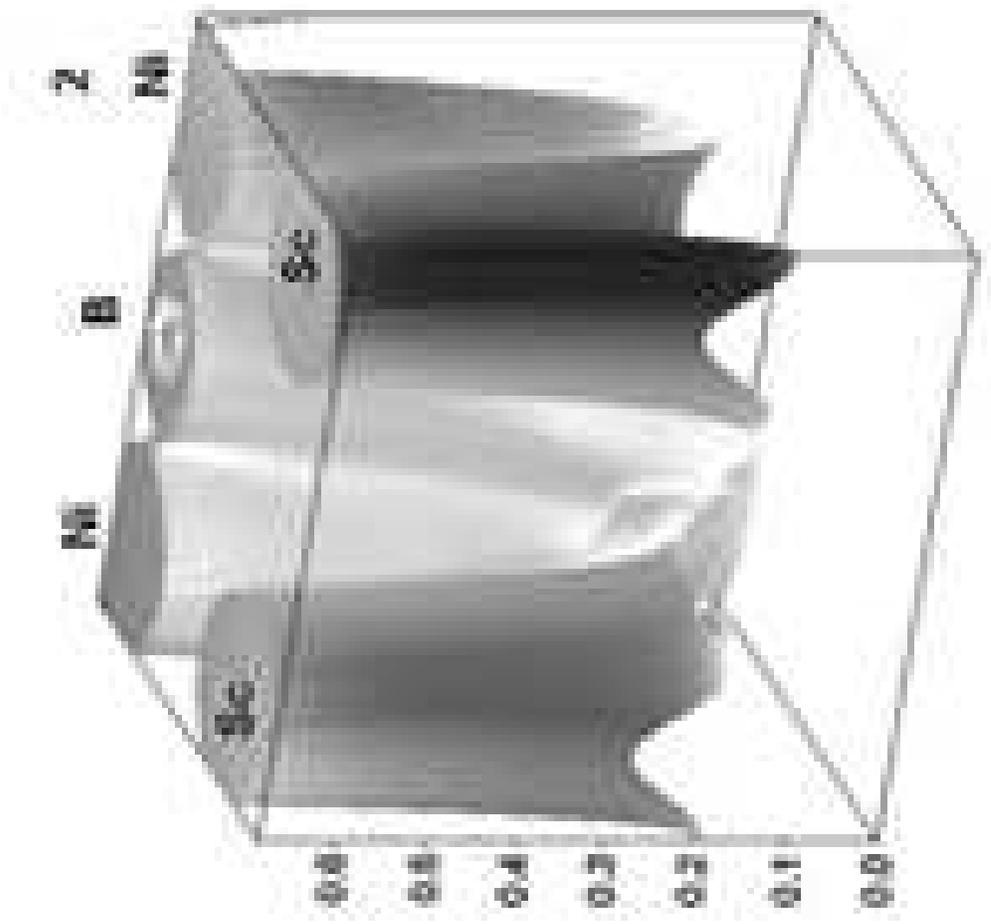
\end{document}